\documentclass{article}

\usepackage{amsmath,amsfonts,amssymb,bm}
\usepackage{mathptmx}
\usepackage{newtxtext}
\usepackage{newtxmath}

\usepackage{natbib}
\usepackage[utf8]{inputenc}
\usepackage{graphicx}
\usepackage[french,english]{babel}
\usepackage[T1]{fontenc}
\usepackage{authblk}
\usepackage{url}

\title{Surrogate Models for Rainfall Nowcasting}

\begin{document}

    \author[4]{Naty Citlali Cabrera-Guti\'errez \thanks{Corresponding author: \texttt{cabrera@cerfacs.fr}}}
    \author[4]{Hadrien Godé}
    \author[4]{Jean-Christophe Jouhaud}
    \author[1]{Mohamed Chafik Bakkay}
    \author[1]{Valentin Kivachuk Burd\'{a}}
    \author[1]{Florian Dupuy}
    \author[1]{Maud-Alix Mader}
    \author[2]{Olivier Mestre}
    \author[1]{Guillaume Oller}
    \author[3]{Mathieu Serrurier}
    \author[2]{Michaël Zamo}

    \affil[1]{Institut de Recherche Technologique Saint-Exupéry, Toulouse, France}
    \affil[2]{Météo-France, Direction des Opérations pour la Production, 42 avenue Gaspard Coriolis, 31057 Toulouse cedex 07, France and CNRM/GAME, Météo-France/CNRS URA 1357, Toulouse, France}
    \affil[3]{IRIT, Université Paul Sabatier, Toulouse, France}
    \affil[4]{CERFACS, Toulouse, France}

\maketitle

\begin{abstract}
Nowcasting (or short-term weather forecasting) is particularly important in the case of extreme events as it helps prevent human losses. Many of our activities, however, also depend on the weather. Therefore, nowcasting has shown to be useful in many different domains. Currently, immediate rainfall forecasts in France are calculated using the Arome-NWC  model developed by Météo-France, which is a complex physical model. Arome-NWC forecasts are stored with a 15 minute time interval. A higher time resolution is, however, desirable for other meteorological applications. Complex model calculations, such as Arome-NWC, can be very expensive and time consuming. A surrogate model aims at producing results which are very close to the ones obtained using a complex model, but with largely reduced calculation times. Building a surrogate model requires only a few calculations with the real model. Once the surrogate model is built, further calculations can be quickly realized. In this study, we propose to build surrogate models for immediate rainfall forecasts with two different approaches: combining Proper Orthogonal Decomposition (POD) and Kriging, or combining POD and Random Forest (RF). We show that results obtained with our surrogate models are not only close to the ones obtained by Arome-NWC, but they also have a higher time resolution (1 minute) with a reduced calculation time.
\end{abstract}


\section{Introduction}

Arome-NWC  is a fine grid (1.3 km) nowcasting system that is based on an existing Arome mesoscale model, from which background and lateral boundary conditions are taken. Arome, in turn, uses the non-hydrostatic version of the Euler equations and takes its lateral boundary conditions from Arpege (Météo-France's global model). Arome-NWC and Arome  share the same physics, the same dynamics, the same 3DVar assimilation \cite{Merlet2017}.

Arome-NWC  has a maximum forecast range of 6 hours. Every hour $H$, the last available Arome-NWC forecast and observations in the interval $[H-10,H+10[ \,$ are used to give rainfall forecasts with a temporal resolution of 1 minute, but, due to storage reasons, they are only saved every 15 minutes. Therefore, rainfall forecasts are available every 15 minutes, from $H+15$ min to $H+6$ hours and the output is available at $H+30$ min \cite{LAuger2015}. Making rainfall forecasts with a higher time resolution available would be useful for operational applications in nowcasting. 

In this study, we propose to build surrogate models to give rainfall forecasts with a higher time resolution and a reduced calculation time compared to the time needed to run the Arome-NWC  model. In particular, our surrogate models are built by combining two different techniques: Proper Orthogonal Decomposition (POD), which carries out dimensionality reduction while keeping existing coherent structures \cite{JMBeckers2003} \cite{DAndrea2001}, and a regression technique, in our case Kriging (also known as Gaussian Process Regression) or Random Forest. POD and Kriging have been successfully used together for spatial interpolation \cite{GBiau1999} and to build surrogate models in aerodynamics \cite{RDupuis2018}, in combustion systems \cite{Aversano2019}, or in geology \cite{Lange2010}. POD and Random Forest have been used in genetics \cite{FBertolini2015} and for human activity recognition \cite{SBalli2019}, among others.

Kriging \cite{Duraiswami2010EfficientKF} \cite{Kilibarda2014} and Random Forest \cite{Kilibarda2014} can be used for spatio-temporal interpolations. However, spatio-temporal interpolations with Kriging have a high computational cost. In our case, Arome-NWC already has a high spatial resolution, so a spatial interpolation was not needed. Therefore we focused only on doing a time interpolation.

This article is structured as follows: First, in section \ref{building_SM_section} we explain the theory of the three techniques used here to build a surrogate model: Proper Orthogonal Decomposition (\ref{POD_theory}), Kriging (\ref{Kriging_theory}) and Random Forest (\ref{RF_theory}). In section \ref{Rainfall_FC}
we show that Kriging and Random Forest can be used on the POD basis to build surrogate models and do time interpolations of rainfall forecasts. In subsection \ref{previous_data_section} the use of extra data to build a Kriging-based surrogate model is explored. Our surrogate models are evaluated in section \ref{evaluation_section} and some particular cases are analyzed in section \ref{analysis_section}. Finally, a general discussion and our conclusions can be found in section \ref{discussion_conclusion_section}.

\section{Building a surrogate model} \label{building_SM_section}

In this section we briefly introduce Proper Orthogonal Decomposition, as well as the two regression methods (Kriging and Random Forest) used to build our surrogate models.

\subsection{Proper Orthogonal Decomposition (POD)} \label{POD_theory}

Proper Orthogonal Decomposition (also known as PCA: Principal Component Analysis or EOF: Empirical Orthogonal Functions) is a wide-spread technique used for dimensionality reduction. POD takes data in $n$ dimensions and finds a $k$-dimensional space (where $k<n$) which captures the maximum amount of variance and in which the projection of these data has a minimized projection error.  

A n-dimensional variable $f$ can be written as the sum of its mean value and a residual:

\begin{equation}
f_{i}=f(\boldsymbol{x}, t_{i}) = \Bar{f}  + V(\boldsymbol{x},t_{i})   \text{ for } i=1,..m
\end{equation}

\noindent
here, $\Bar{f}$ corresponds to the mean value and $V_{i}=V(\boldsymbol{x},t_{i})$ to the residual.

POD maximizes the projection of $V_{i}$ (note that $V_{i}=f_{i}-\Bar{f}$) on an orthonormal basis $\boldsymbol{\phi_{i}}$. That is:

\begin{equation}
 \underset{\phi_{1},.. \phi_{n}}{max} \overset{m}{ \underset{i=1}{\sum}} \overset{n}{ \underset{j=1}{\sum}}  \rvert (V_{i} \cdot \boldsymbol{\phi_{j}} )  \rvert^{2}
\end{equation}

\noindent
where $(\boldsymbol{\phi_{i}} \cdot \boldsymbol{\phi_{j}} ) = \delta_{i,j}$, with $ \delta_{i,j}$ the Kronecker delta ($ \delta_{i,j}=1$ if $i=j$ and $\delta_{i,j}=0$ otherwise ).

The total variation in the data is given by:
\begin{equation}
\frac{1}{m} \overset{m}{ \underset{i=1}{\sum}} \rvert \rvert  V_{i} \rvert \rvert^{2}
\end{equation}

Typically, $k$ is chosen to be the smallest value that satisfies:

\begin{equation}
\frac{\frac{1}{m} \overset{m}{ \underset{i=1}{\sum}} \rvert \rvert  V_{i} - \hat{V_{i}} \rvert \rvert^{2}}{\frac{1}{m} \overset{m}{ \underset{i=1}{\sum}} \rvert \rvert  V_{i} \rvert   \rvert^{2}} \leq 0.01
\end{equation}

\noindent
with 

\begin{equation}
\hat{V_{i}} =  \overset{k}{ \underset{i=1}{\sum}} a_{i} \boldsymbol{\phi_{i}} 
\end{equation}

\noindent
where $\hat{V_{i}}$ is the approximation done on the POD basis keeping only the main $k$ modes of the basis. It is then said that $99\%$ of the variance (or energy) is retained.

\subsection{Kriging} \label{Kriging_theory}

Kriging is an interpolation technique that uses known values of residuals $V$ to calculate semivariances and covariances and build a surrogate model, which corresponds to the best linear unbiased predictor (that is, with the minimum expected mean-squared error).  This predictor (surrogate model) can then be used to estimate the value of residuals at other (unknown) positions (in time and/or space). 

For this technique, it is assumed that the variance of $V$ is constant and the covariance of $V$ at two different points only depends on their distance (in time and/or space).

\noindent
A new prediction is then given by:

\begin{equation}
\hat{V}^{Kr}_{0} =  \boldsymbol{c}^{-1}\boldsymbol{c}_{0} \boldsymbol{V}
\end{equation}

where $\boldsymbol{c}$ is the variance-covariance matrix of the known residuals, $\boldsymbol{c}_{0}$ is a vector of covariances between the known and the predicted residuals and $\boldsymbol{V}$ is the vector of known residuals.

In our case, we used POD for dimensionality reduction and Kriging is done on the POD basis. This means that: 

\begin{equation}
\hat{V}^{Kr}_{0} =  \overset{k}{ \underset{i=1}{\sum}} a_{i} \boldsymbol{\phi_{i}} 
\end{equation}

\noindent
where $a= \boldsymbol{c'}^{-1}\boldsymbol{c'}_{0}$. So for each new prediction, we only need to calculate the value of the coefficient $a_{i}$ for each POD mode $\boldsymbol{\phi_{i}}$. In our case, we interpolate the rainfall value, which can be seen as a rainfall map, at unknown times.

For more details on this technique, the interested reader is referred to \cite{Kige1951} \cite{Matheron1971} \cite{Rasmussen2006}.

\subsection{Random Forest} \label{RF_theory}
Random Forest is a method that uses a combination of decision tree predictors and makes a prediction by averaging the prediction of each component tree. Individually, those predictors might be far from the expected results, but combined, they converge to a good prediction. This is known as "wisdom of crowd". 

First, features (variables) are randomly selected and used to create decision trees which are extended using an optimized version of CART, an algorithm to decompose the space \cite{LBreiman1984}. This is done as follows: a threshold value, that splits the samples into two subspaces, is selected in such a way that the quadratic error using all samples is minimized. The threshold is then known as a node. The process is recursively repeated in each subspace, leading to child nodes. At the end, each subspace is known as a node leaf and each leaf determines the value of the prediction. If the depth of the trees (the number of splits before coming to a prediction) is not limited, Random Forest becomes an exact predictor. That is, the prediction for a sample used to create the model is the same as the originally given value.

A variant of Random Forest is used here: Extra trees \cite{PGeurts2006}, also known as Extremely Randomized Trees. The main difference is that instead of computing the locally optimal feature-split combination, for each feature a random value is selected as the threshold for the split. The best of these randomly generated thresholds is picked as the splitting rule. This means that, both, the subset of features and the splits are chosen based on random values.

To make a prediction, Random Forest tests the new sample on each tree. The final prediction is the mean value of each tree prediction \cite{LBreiman2001}. The main benefit of Random Forest is the speed. The model and the predictions are made in a very short time. Besides Kriging, we used Random Forest to obtain a prediction in the POD space.

\section{Application to rainfall forecasting} \label{Rainfall_FC}
As previously stated, every hour, Arome-NWC  outputs 24 predictions (a prediction every 15 minutes for 6 hours), which constitutes the total number of samples that can be used to build our surrogate model. We built surrogate models using different numbers of samples: 5, 7, 9 and 13. This allowed us to take the 'unused' samples as reference points to evaluate our surrogate models. It is important to notice that we use a regression method to carry out an interpolation and therefore the first ($H + 15$ min) and the last ($H+ 6$ hours) samples must be included to build up the surrogate model.

A key point is that instead of considering a scalar rainfall value for each latitude and longitude location at a given time, we consider rainfall as a vector of size $(1,n_{lat}* n_{lon})$ (where $n_{lat}$ and $n_{lon}$ are the number of latitude and longitude points, respectively), that is, a vector containing all the rainfall values over our region of interest. Since we know the latitude and longitude ranges of our data, once the interpolation is done, we can restructure our data in their original shape and identify the corresponding latitude and longitude values. This has the advantage of reducing the amount of memory required for the calculations and the calculation time itself.

All calculations shown here were done using batman \cite{batman_paper}, a python tool developed at Cerfacs based on the scikit-learn library. Note that in our case, a sample corresponds to a time and its corresponding rainfall vector (of size $ (1,n_{lat}* n_{lon})$). In this work, "forecasts" or "simulations" refer to Arome-NWC data, while "predictions" or "time interpolations" refer to the output of our surrogate models. Our surrogate models were evaluated by comparing their predictions to the original Arome-NWC  simulation (unused samples). We calculated the $Q2$ coefficient, which represents the percentage of variance explained by our model. Typically, $Q2 \approx 0.8$ indicates that the created surrogate model is a good approximation of the original complex model.

\begin{figure}[t]
\includegraphics[width=\linewidth]{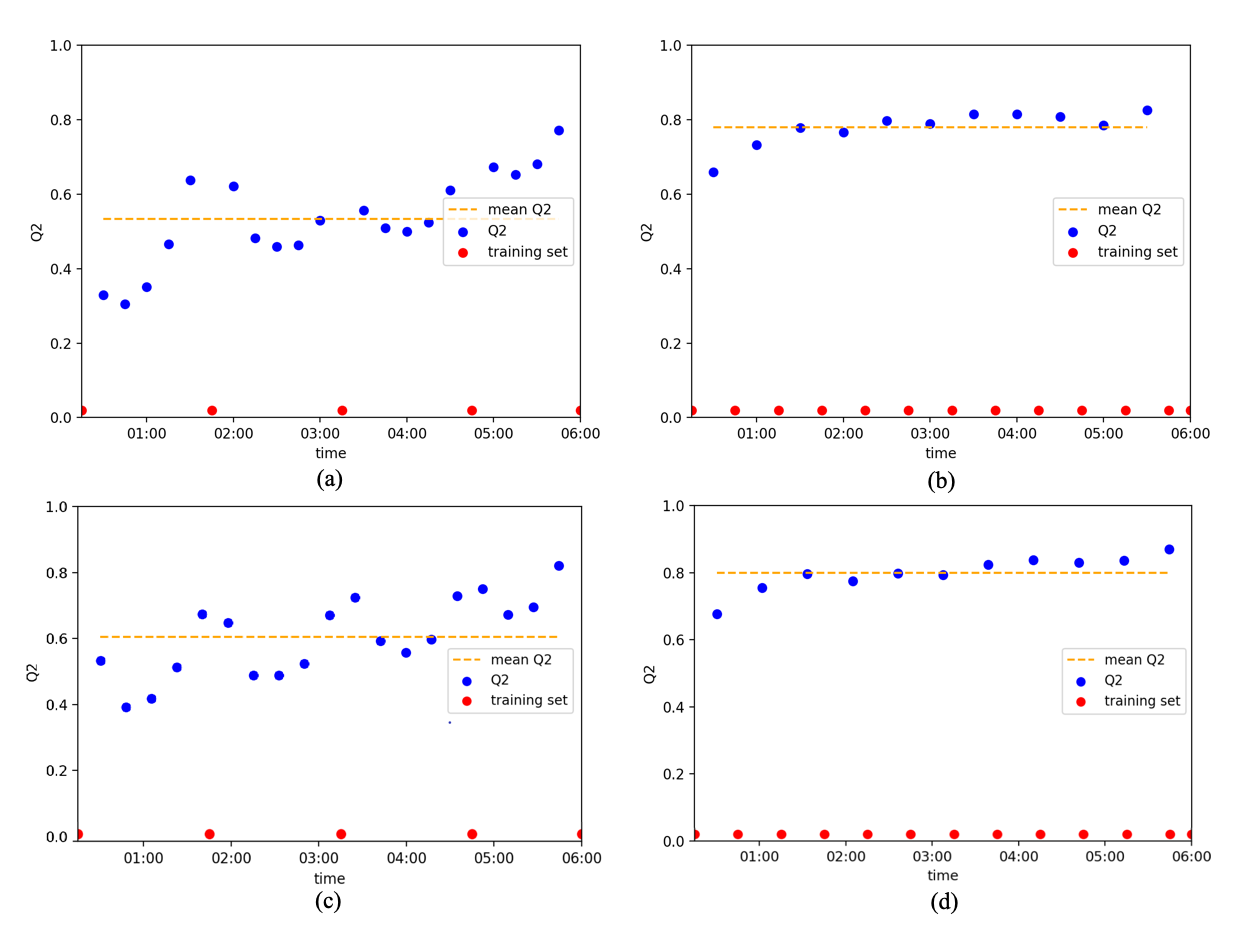}
\caption{Evolution of $Q2$ when using 5 and 13 training samples to create our surrogate model with Kriging ((a) and (b)) and with Random Forest ((c) and (d)). Blue points correspond to individual $Q2$ values. Red points indicate the samples that were chosen to create the surrogate model. The orange dashed line corresponds to the mean value of $Q2$.}
\label{fig: q2_diff_samples_kriging}
\end{figure}

Figures \ref{fig: q2_diff_samples_kriging}(a) and \ref{fig: q2_diff_samples_kriging}(b) (\ref{fig: q2_diff_samples_kriging}(c) and \ref{fig: q2_diff_samples_kriging}(d)) show the evolution of $Q2$ when using 5 and 13 training samples and Kriging  (Random Forest) to create the surrogate model for the Arome-NWC simulation of January $4^{th}$, 2018 at 0:00 A. M.\footnote{Here we considered a region of latitudes from $42$ to $54$ degrees N and longitudes from $6$ degrees W to $8$ degrees E.}. As a prediction gets closer to a training sample, its $Q2$ increases. Also, as we go further in time, predictions tend to have higher $Q2$ values, which indicates that points preceding them are important. As expected, the value of $Q2$ increases when more training samples are used to build the model and it goes from a mean value (see orange dashed lines) of $Q2\approx 0.5$  ($Q2\approx 0.6$) for 5 training samples to a mean value of $Q2\approx 0.8$  ($Q2\approx 0.8$) for 13 training samples using Kriging (Random Forest). Therefore, the value of $Q2$ is expected to increase even further if all 24 available training samples are used. However, when all 24 training samples are used, the value of $Q2$ not be calculated. 
A more complete analysis will be presented in section \ref{evaluation_section}

\subsection*{Using previous data} \label{previous_data_section}

So far we've shown that surrogate models can be used to build good time interpolations. However, it should be kept in mind, that the $Q2$ value will change from one case to another. In some cases, the predictions obtained with our surrogate model will have lower $Q2$ values. One of the limiting factors when creating our surrogate model is the number of available training samples. In order to increase the number of training samples, three different options have been explored:  a) using previous Arome-NWC data b) using previous radar data and c) using previous Antilope data \footnote{For this part, a smaller region was considered to exclude areas where radar and antilope data might be missing. Latitudes go from $44$ to $49$ degrees N, while longitudes go from $0$ to $8$ degrees E}. 

\begin{figure}[t]
\includegraphics[width=\linewidth]{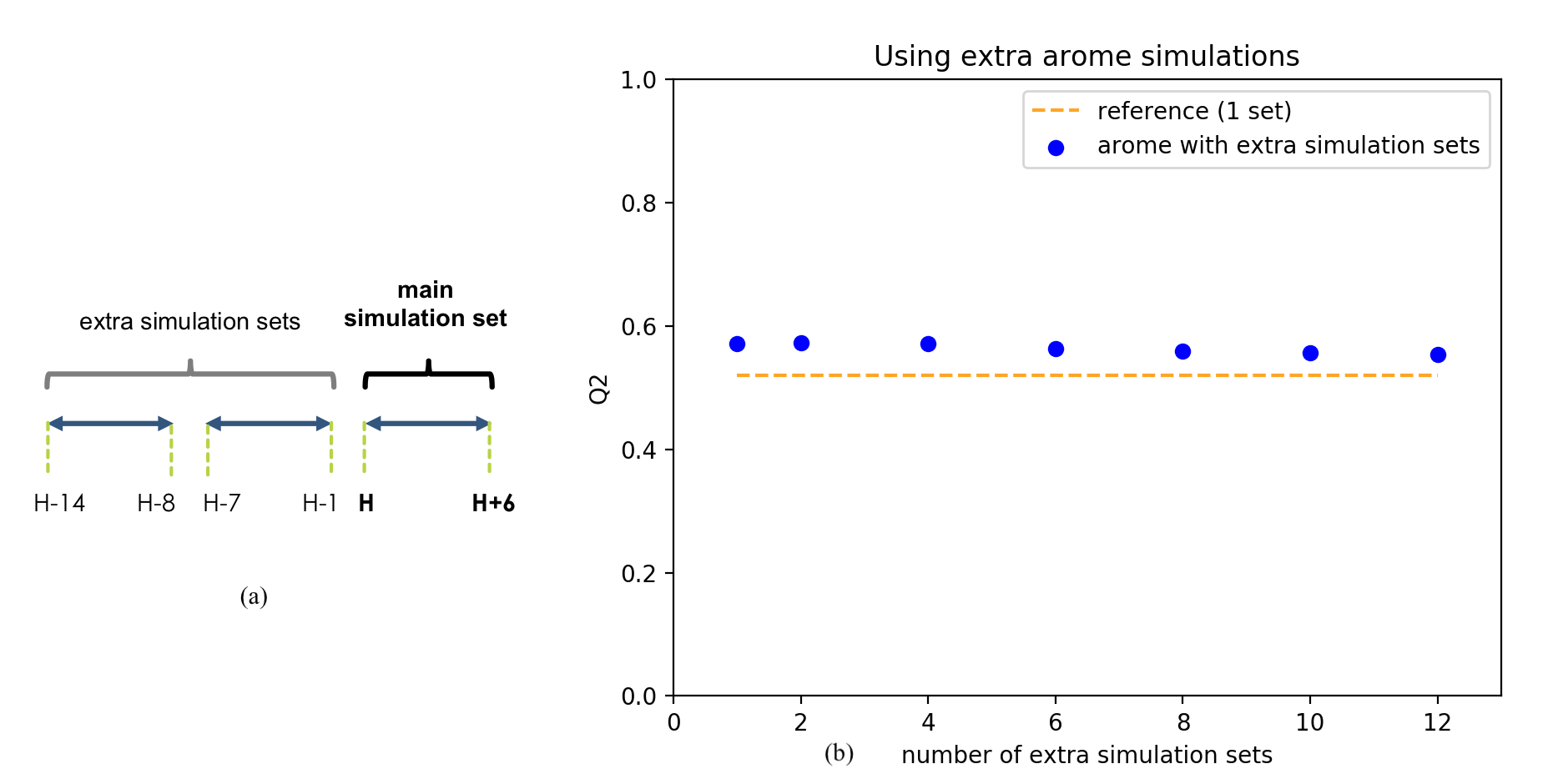}
\caption{(a) Scheme of Arome-NWC simulation sets that can be used as extra data to build a Kriging-based surrogate model (b) evolution of $Q2$ with different numbers of extra Arome-NWC simulation sets (and 13 samples from the main simulation set) are used to build a Kriging-based surrogate model. The orange dashed line corresponds to the case where a single AromeNWC simulation set is used and serves as a reference value. Data from January 16$^{th}$, 2018  at  $0$ A.M.}
\label{fig: q2_General_analysis_eA}
\end{figure}

We can use Arome-NWC data from other rainfall forecasts calculated at previous times. In this case, two points should be kept in mind. First, simulations should be chosen in such a way, that there is only one forecast at a given time. Since Arome-NWC returns forecasts for 6 hours, the first extra set of simulations that could be used would be the one from $H-6$. However, since every time Arome-NWC is launched it uses different initial conditions, there would be no continuity if simulation sets $H-6$ and $H$ were to be used. One way to get around this, is to also break the continuity in time. Therefore, only sets of simulations with one hour of difference are used (see figure \ref{fig: q2_General_analysis_eA}(a)).

All results shown in this section were obtained using Kriging as the regression method to build the surrogate model. Data from January 16$^{th}$, 2018  at  $0$ A.M. were used here.  As can be seen in figure \ref{fig: q2_General_analysis_eA}(b) using more Arome-NWC simulation sets has a positive effect on the value of $Q2$ when Kriging is used to build a surrogate model. However, as the number of Arome-NWC simulation sets is greater than $4$, the value of $Q2$ gets closer to the reference value. Therefore, we decided to use only two extra Arome-NWC simulation sets (3 simulation sets in total) and compare those results with the other two cases (using radar/Antilope data).

Radar data correspond to rainfall measurements done by radars. We use here 2-D data, but a 3-D version of these data is fed to Arome-NWC  to set the initial conditions. Therefore, all data previous to Arome-NWC simulations can be used and preserve time continuity. Radar measurements are available every 5 minutes.

Antilope data correspond to a reanalysis which combines radar data and rainfall observations at ground level. Therefore, also in this case all data previous to Arome-NWC simulations can be used and preserve time continuity. Antilope data are available every 15 minutes. There are two versions of Antilope, we use here the "real time" output that has a calculation time of $9$ minutes.

\begin{figure}[t]
\includegraphics[width=\linewidth]{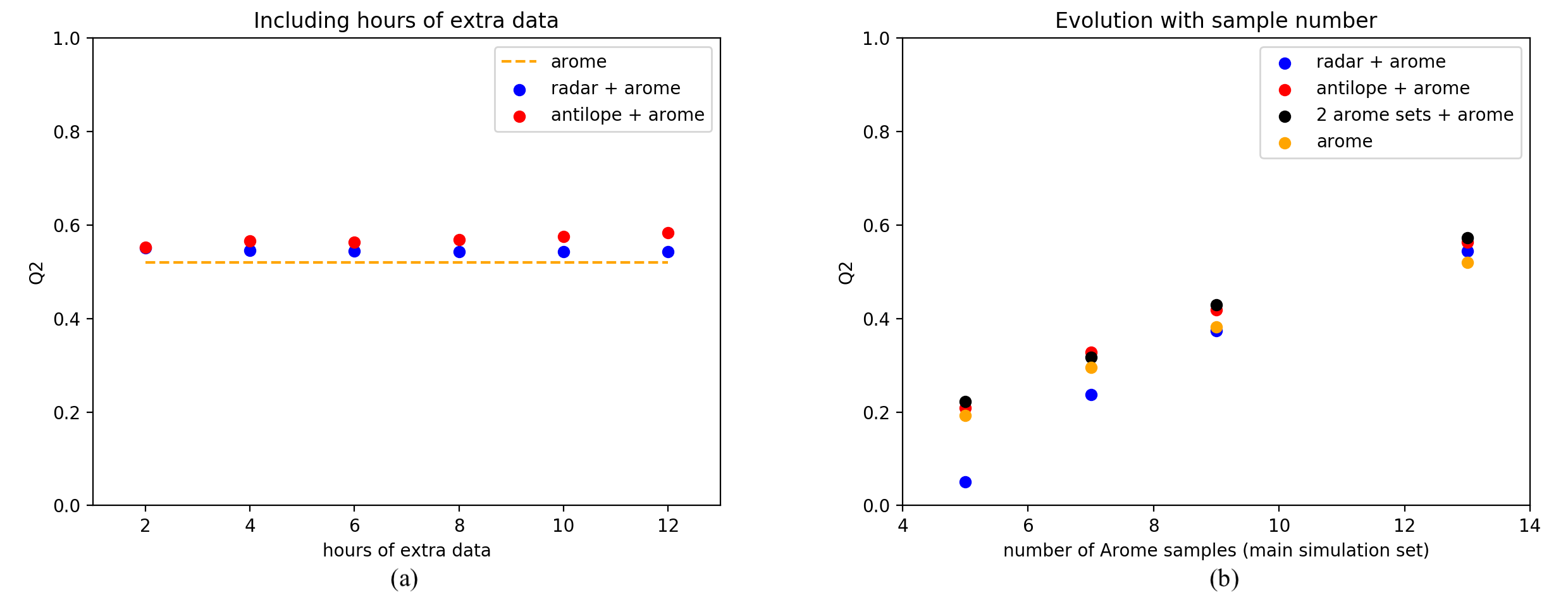}
\caption{Evolution of $Q2$ for (a) different numbers of hours of radar and Antilope data and 13 Arome-NWC samples (b) different numbers of training samples from the main Arome-NWC simulation set. Three different kinds of extra data are compared: previous radar data (6h, blue dots), previous Antilope data (6h, red dots) and 2 previous Arome-NWC simulation sets (black dots). The case of Arome-NWC with no extra data is also shown (orange dots). Data from January 16$^{th}$, 2018  at  $00:00$ a.m.  }
\label{fig: q2_General_analysis_hours_samples}
\end{figure}

Figure \ref{fig: q2_General_analysis_hours_samples}(a) shows the evolution of $Q2$ when using different numbers of hours of previous radar (black dots) and Antilope (red dots) data together with 13 Arome-NWC samples to build a Kriging-based surrogate model. While both increase the value of $Q2$ compared to the case where a single Arome-NWC simulation is used (orange dashed line), using Antilope data leads to a better result. However, Antilope is a reanalysis of radar data and for this reason, all measured data need to be processed to create the corresponding Antilope data. Therefore, Antilope data are not immediately available. Moreover, the value of $Q2$ obtained when using 10 hours of extra Antilope data (and 13 Arome-NWC samples) is roughly the same we obtained when using two extra Arome-NWC simulation sets and 13 samples from the main Arome-NWC simulation set (3 Arome-NWC simulation sets in total).

The effect of using different numbers of training samples from the main Arome-NWC simulation set to build a Kriging-based surrogate model is shown in figure \ref{fig: q2_General_analysis_hours_samples}(b) for all three cases: 6 hours of extra radar data (blue dots), 6 hours of extra Antilope data (red dots) and 2 extra Arome-NWC simulation sets (black points). The case where a single Arome-NWC simulation set is used (no extra data) is also shown (orange points). As expected, in all cases the value of $Q2$ increases with the number of training samples. For 13 training samples (from the main Arome-NWC simulation set), $Q2\approx0.6$, so we expect it to be even higher when using all 24 samples from the main Arome-NWC simulation set (this value, however, can't be calculated). We see that using 2 extra Arome-NWC simulation sets performs slightly better than using 6 hours of Antilope data (or extra radar data), hence this solution has been chosen to build Kriging-based surrogate models.

\section{Evaluation of predictions} \label{evaluation_section}

We chose data from 10 different dates in 2018\footnote{January $3^{rd}$,$4^{th}$, $13^{th}$,$14^{th}$,$16^{th}$ and February $3^{rd}$,$4^{th}$,$13^{th}$,$14^{th}$,$16^{th}$. The Arome-NWC simulation from 0 A.M. was used.} to study the evolution of $Q2$ when using different numbers of samples to build a surrogate model. Here, we also consider the case where extra simulation sets are used or not.

As explained before, when using extra Arome-NWC simulations we have to make sure that there is only one forecast at a given time. Since initial conditions are updated every hour and Arome-NWC gives a forecast for the following 6 hours, only simulations with one hour of difference are used to build Kriging-based surrogate models (there is no continuity in the initial conditions, so we also break the continuity in time, see figure \ref{fig: q2_General_analysis_eA}(a)).

In the case of Random Forest, a rainfall forecast at a given time is considered as a sample that has no particular connection with other samples and because there is no physical background, having continuity in the initial conditions or not, doesn't make any difference. Therefore, two consecutive Arome-NWC simulation sets $H-6$ and $H$ (the main simulation set and one extra simulation set) are used in this case.

First, we considered the case where no extra simulation sets are used and calculated the value of $Q2$ when using $5$, $7$, $9$,$13$ and $23$ samples to build the surrogate model.

\begin{figure}[t]
\includegraphics[width=\linewidth]{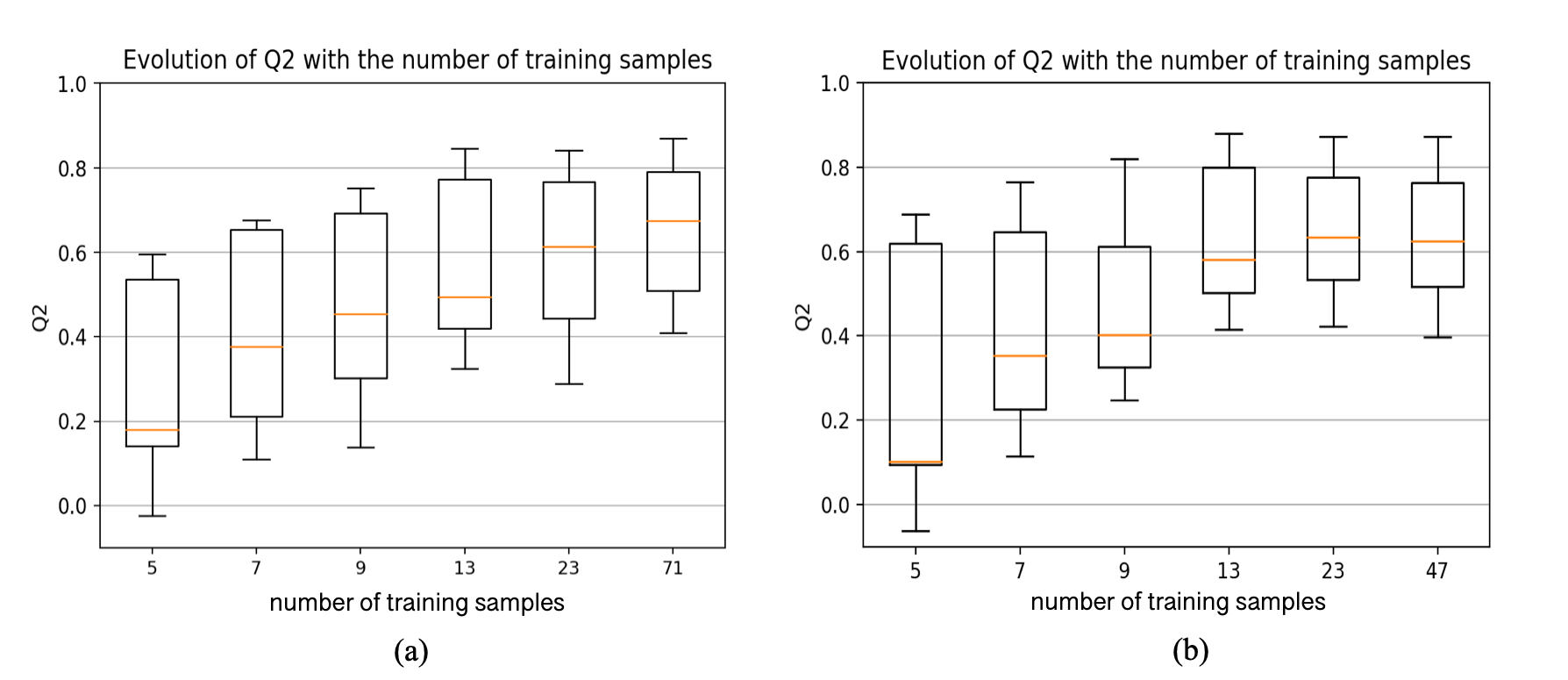}
\caption{(a) Kriging: Evolution of $Q2$ with (71 samples) and without extra Arome-NWC simulation sets (5, 7, 9, 13 and 23 samples) (b) Random Forest: Evolution of $Q2$ with (47 samples) and without an extra Arome-NWC simulation set (5, 7, 9, 13 and 23 samples). The orange line represents the median, while the bottom and top of the squares represent the first and third quartiles. The black horizontal lines represent the minimum (bottom) and maximum (top) values. Points outside of this area represent outliers.}
\label{fig: q2_General_analysis}
\end{figure}

Then, we included the use of extra simulation sets. In the case of Kriging, we used 3 simulation sets (2 extra simulation sets and the main simulation set), as for Random Forest, two contiguous simulation sets are used. We then calculated the value of $Q2$ when using $71$ ($47$) samples to build a surrogate model with Kriging (Random Forest). As already stated, when all available samples are used to build the surrogate model, it isn't possible to calculate the value of $Q2$. One can, however, get an estimate of this value using Leave One Out (LOO) cross validation: one sample is left out from the training set used to build the surrogate model. The quality of the built surrogate model is then evaluated on the one sample that was left out. Usually, this is done as many times as samples there are, each time leaving a different sample out, and the mean value is then calculated. The evaluation obtained using LOO underestimates the real $Q2$ value of the model, but gives an estimate of its performance. In our case, LOO was repeated only 5 times (leaving out the samples corresponding to $H+0:30$, $H+1:30$, $H+2:30$ , $H+3:30$, $H+4:30$, $H+5:30$). Leave One Out cross validation was used (making sure the sample that was left out belonged to the main simulation set) to evaluate our surrogate models when using $23$ and $47$ (Random Forest) or $71$ (Kriging) samples. When $5$, $7$, $9$ and $13$ samples are used to build the surrogate model, all the other ("unused") samples are used to calculate the value of $Q2$. Figure \ref{fig: q2_General_analysis} summarizes the results that we obtained.

In the case of Random Forest, the best results are obtained when $23$ samples are used, with a mean value of $Q2\approx 0.6$. It is interesting to notice that, in this case, using one more Arome-NWC simulation set has a negative effect on the value of $Q2$ (see figure \ref{fig: q2_General_analysis}(b)). Indeed, Random Forest doesn't learn any physics from the data, so no connection is seen between two samples that are contiguous in time. Furthermore, if the data used to build the surrogate model are distant (in time) from the prediction time, they have no real utility to make the prediction as the corresponding nodes in the decision tree remain unused.

In the case of Kriging, however, using more Arome-NWC simulation sets has a positive effect on the value of $Q2$ (see figure \ref{fig: q2_General_analysis}(a)). The best results are obtained when $71$ samples are used, where a mean value of $Q2 \approx 0.7$ is calculated . In fact, when more Arome-NWC simulation sets are used, the value of $Q2$ improves for nearly all predictions, and the improvement is quite significant for the first time values for which we had just a few samples (preceding them) before adding extra Arome-NWC simulation sets. This confirms that, contrary to surrogate models built using Random Forest, the physics of the problem is preserved.

Météo-France kindly provided us with Arome-NWC data from 2019 with more samples than usually stored. That is, for every simulation from $H+15$ minutes to $H+6$ hours, we had a rainfall forecast every $5$ minutes. While these data are not usually available (operationally, rainfall forecasts are only stored every 15 minutes), they were useful to calculate the real value of $Q2$. In order to do that, we used the $24$ samples (one every $15$ minutes) that are usually stored for each Arome-NWC simulation and built surrogate models. The value of $Q2$ corresponding to those surrogate models was then calculated using all the samples that were not used to build the model. This was done for $5$ different simulations \footnote{Data used: August $9^{th}$ at $10$ A.M. and $10$ P.M.,  October $22^{th}$ at $12$ P.M., November $4^{th}$ at $3$ P.M. and November $24^{th}$ at $0$ A.M.} and we found a mean $Q2$ value of $0.89$ ($0.92$) for surrogate models built using Kriging (Random Forest). This confirms that that both regression methods (Kriging and Random Forest) combined with POD lead to good results and their quality is roughly equivalent.

\section{Analysis of results} \label{analysis_section}

Our two different approaches were tested on different data corresponding to different meteorological situations. In the following, we show different results obtained with surrogate models built with 23 samples from the main Arome-NWC simulation set \footnote{The region studied here corresponds to latitudes from $42$ to $51.5$ degrees N and longitudes from $6$ degrees W to $8$ degrees E. For all cases shown here the time of main Arome-NWC simulation set is 0 A.M.}. In the case of Kriging, 2 extra Arome-NWC simulation sets were used whereas in the case of Random Forest only the main simulation set was used. In practice, all 24 samples from the main simulation set must be used. Nevertheless, using 23 samples is useful for comparison purposes. The values of $Q2$ in this section were calculated using the one sample that was left out while building our surrogate models.

\begin{figure}[t]
\includegraphics[width=\linewidth]{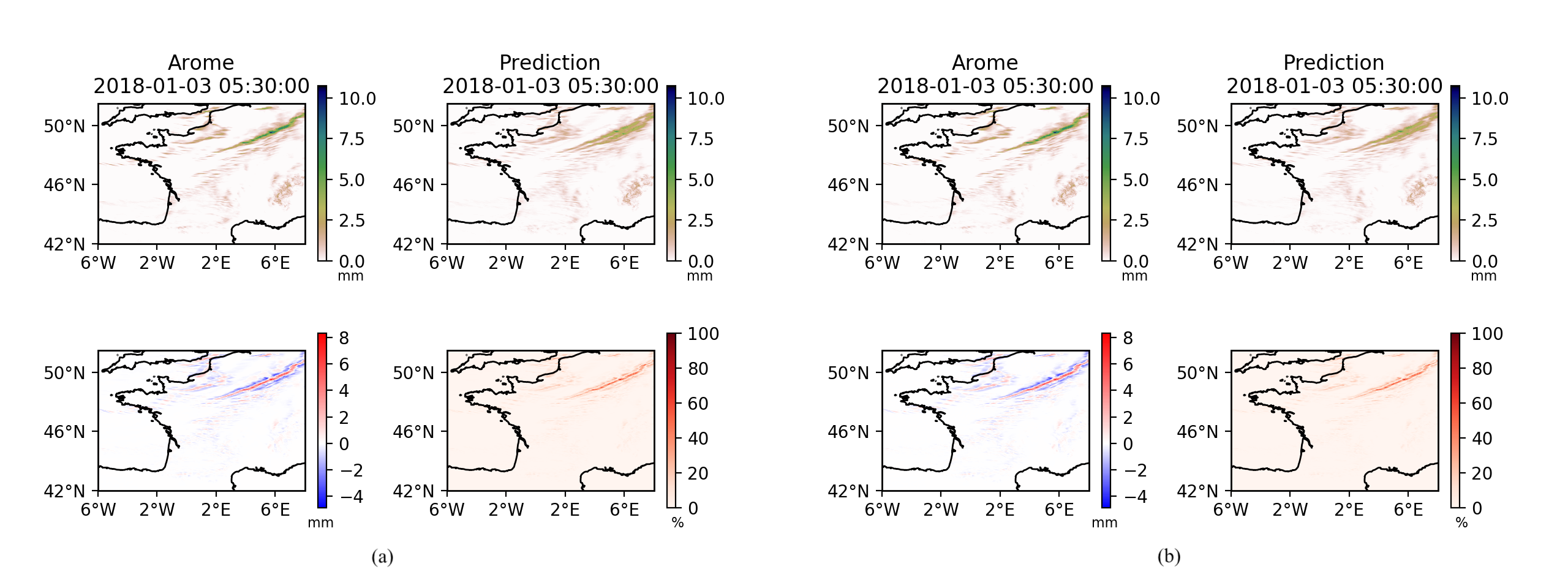}
\caption{(a) Kriging prediction. Surrogate model built using 23 samples from the main simulation set and 2 extra simulation sets. (b) Random Forest prediction. Surrogate model built using 23 samples from the main simulation set (and no extra simulation sets). For each case: Top left: Arome-NWC forecast for January $3^{rd}$, 2018 at 5:30 A.M. (simulation set from 0 A.M.) Top right: Corresponding surrogate model prediction. Bottom left: Difference of the two of them (Arome-Prediction). Bottom right:  Absolute value of the difference of the two of them divided by the maximal value of the Arome-NWC simulation ($\times 100$). }
\label{fig: Difference20180103}
\end{figure}

 Figure \ref{fig: Difference20180103} shows the results obtained for January the 3$^{rd}$, 2018 at 5:30 A. M. . In this case, rainfall is quickly moving to the south-east. At 3:00 A.M. an extreme event with a rain front becomes clearly visible. Both of our surrogate models, using Kriging or Random Forest, have trouble to predict this rain front (one can almost see two rain fronts pretty close to each other in the prediction) and fail to reproduce the expected rainfall value. In fact, rain fronts are particularly challenging to predict, as they tend to be widened by the surrogate model. In both cases, for Kriging and Random Forest, we obtained a value of $Q2\approx 0.4$

\begin{figure}[t]
\includegraphics[width=\linewidth]{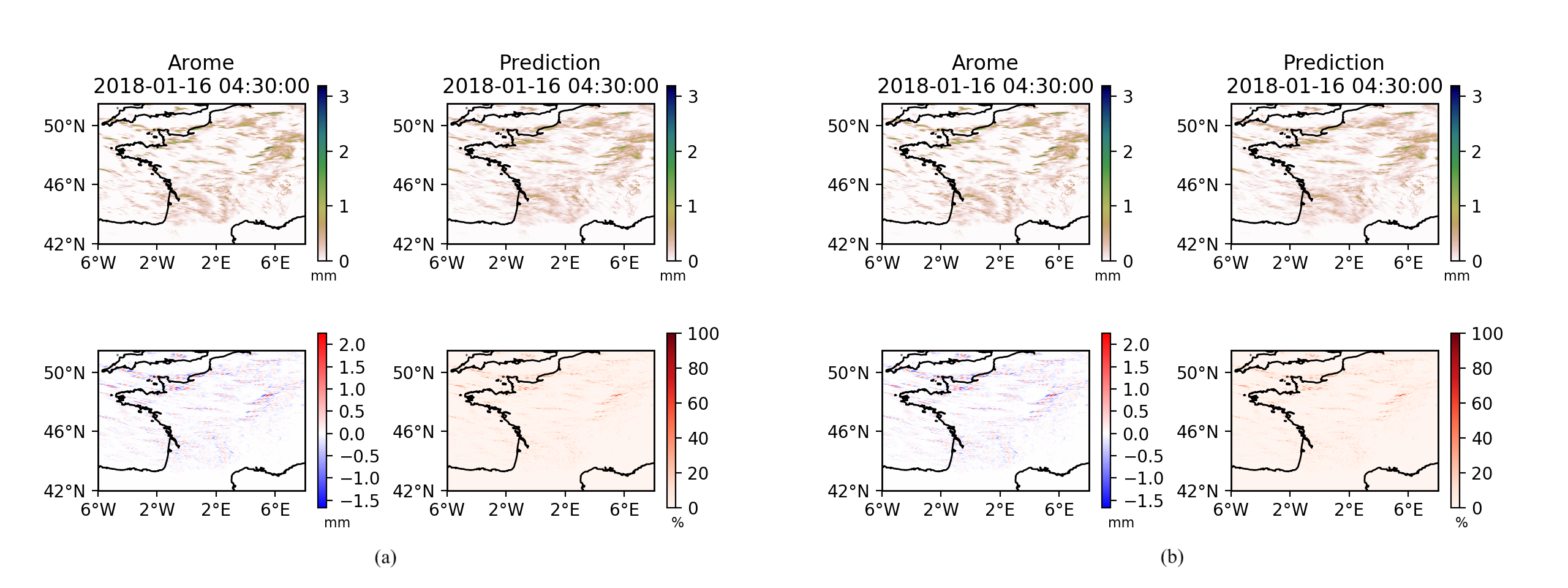}
\caption{(a) Kriging prediction. Surrogate model built using 23 samples from the main simulation set and 2 extra simulation sets. (b) Random Forest prediction. Surrogate model built using 23 samples from the main simulation set (and no extra simulation sets). For each case: Top left: Arome-NWC forecast for January $16^{th}$, 2018 at 4:30 A.M. (simulation set from 0 A.M.) Top right: Corresponding surrogate model prediction. Bottom left: Difference of the two of them (Arome- prediction). Bottom right:  Absolute value of the difference of the two of them divided by the maximal value of the Arome-NWC simulation ($\times 100$). }
\label{fig: Difference20180116}
\end{figure}

Figure \ref{fig: Difference20180116} shows the results obtained for January the 16$^{th}$, 2018 at 4:30 A. M. Here, we observe sparse stratiform rain moving to the east and partially blocked at the highland region of Massif central and the Alps. One can see that in some of the areas where Arome-NWC forecasts no rain, both of our surrogate models predict light rains. A value of $Q2\approx0.51$ and $Q2\approx0.52$ was calculated when using Kriging and Random Forest, respectively.

\begin{figure}[t]
\includegraphics[width=\linewidth]{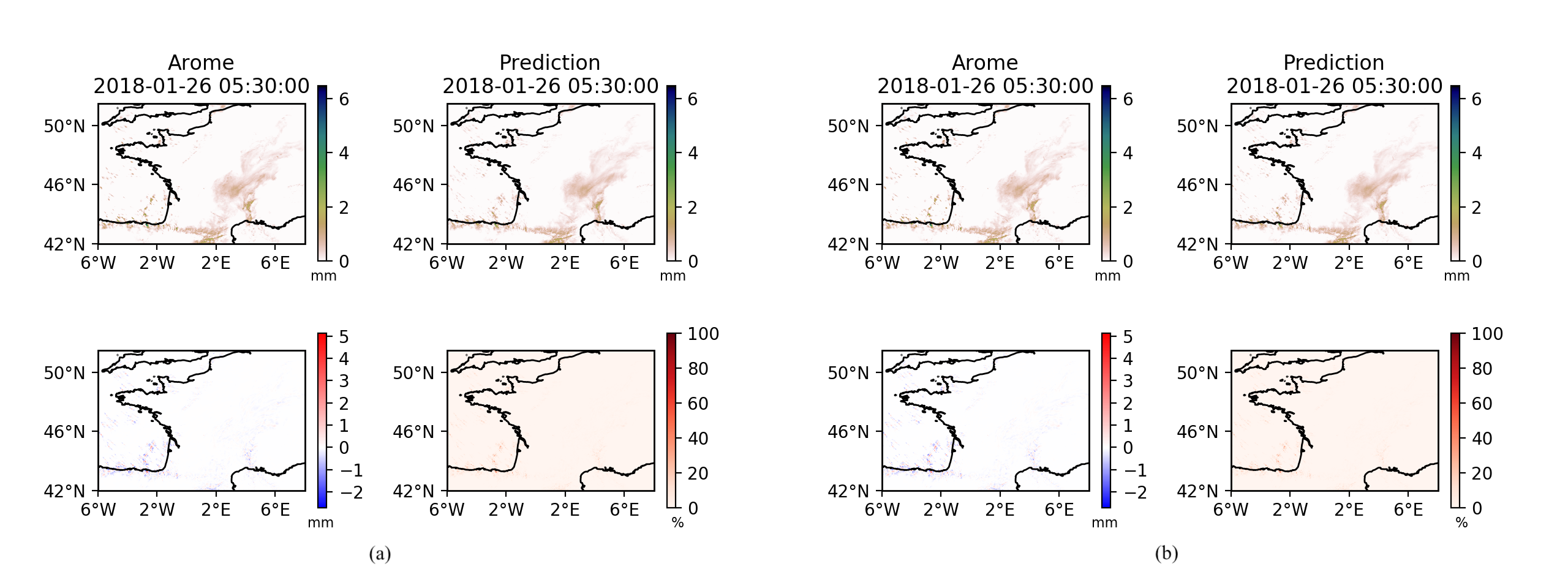}
\caption{(a) Kriging prediction. Surrogate model built using 23 samples from the main simulation set and 2 extra simulation sets. (b) Random Forest prediction. Surrogate model built using 23 samples from the main simulation set (and no extra simulation sets). For each case: Top left: Arome-NWC forecast for January $26^{th}$, 2018 at 5:30 A.M. (simulation set from 0 A.M.) Top right: Corresponding surrogate model prediction. Bottom left: Difference of the two of them (Arome- prediction). Bottom right:  Absolute value of the difference of the two of them divided by the maximal value of the Arome-NWC simulation ($\times 100$). }
\label{fig: Difference20180126}
\end{figure}

Figure \ref{fig: Difference20180126} shows the results obtained for January the 26$^{th}$, 2018 at 5:30 A. M. Here, we observe convective and stratiform rain. Over the Atlantic Ocean, the rain moves south east, while on the western side of France, the rain moves to the north east and is partially blocked at the highlands of Massif central. Here, we found a value of $Q2\approx0.72$ for both of our surrogate models (using Kriging and Random Forest).

\begin{figure}[t]
\includegraphics[width=\linewidth]{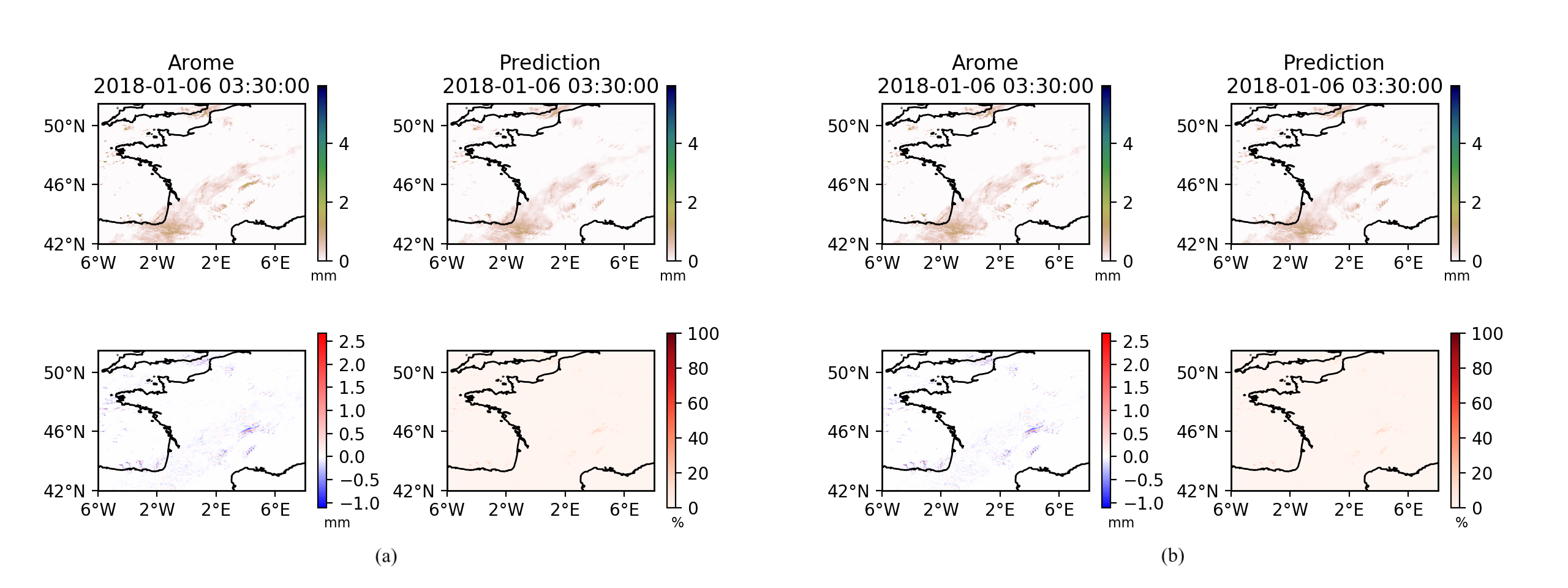}
\caption{(a) Kriging prediction. Surrogate model built using 23 samples from the main simulation set and 2 extra simulation sets. (b) Random Forest prediction. Surrogate model built using 23 samples from the main simulation set (and no extra simulation sets). For each case: Top left: Arome-NWC forecast for January $6^{th}$, 2018 at 3:30 A.M. (simulation set from 0 A.M.) Top right: Corresponding surrogate model prediction. Bottom left: Difference of the two of them (Arome- prediction). Bottom right:  Absolute value of the difference of the two of them divided by the maximal value of the Arome-NWC simulation ($\times 100$). }
\label{fig: Difference20180106}
\end{figure}

Finally, figure \ref{fig: Difference20180106} shows the results for January the 6$^{th}$, 2018 at 3:30 A. M. This case corresponds to convective and stratiform rain moving to the north east, and partially blocked at the highlands of Massif central. In this case, both surrogate models are fairly good, as only small differences are seen when compared to the Arome-NWC forecast. A value of $Q2\approx0.84$ is obtained for both surrogate models (using Kriging and Random Forest).

\section{Discussion and conclusion} \label{discussion_conclusion_section}

In this study, we showed that Kriging and Random Forest combined with POD can be used to build surrogate models for rainfall forecasts, which allow to do reasonably good time interpolations and have a performance that is nearly equivalent. 

Results shown here were obtained using different numbers of samples as some samples needed to be left out to serve as reference and evaluate the surrogate model that was built. Results are improved when using all 24 samples available in the considered Arome-NWC simulation set. In practice, all 24 samples must be used. It should be kept in mind, that the $Q2$ value will change from one case to another. One case that is particularly challenging is predicting rain fronts, as the fronts tend to be widened in the predictions. We also showed that, in the case of Kriging, the physics of the system is preserved and using data from previous Arome-NWC simulations improves the quality of the results. This is however not the case for Random Forest, where using extra Arome-NWC simulation sets can even have a negative effect.

One should be aware of the fact that building a surrogate model in this way has the drawback of working only for the given parameters. This means, that in our case, the predictions will be correct for the considered rainfall regime, but since it evolves with time, the same surrogate model won't give a correct prediction for a different time. Therefore, for each set of Arome-NWC simulations (24 in total: a prediction every 15 minutes for 6 hours), a new surrogate model will have to be built.

All calculations were run on a CPU with 64 GB and 2.5 GHz and took about 19 minutes  (CPU time and equal real time) when using 72 samples (3 Arome-NWC simulations sets in total) and Kriging. In the case of Random Forest, slight modifications were made in the code of batman and we had a CPU time of 44 seconds \footnote{This corresponds to a real time of 28 seconds} (79 seconds) when using 23  samples (71 samples). Using this modified version of batman, the CPU time for Kriging is 109 seconds (96 minutes \footnote{This corresponds to a real time of 300 seconds}) when 23 samples (71 samples) are used to build the surrogate model\footnote{We can however not explain why the calculation time is increased in this case}. Our calculation time includes the calculation of predictions for every minute (not shown), which corresponds to a total of $346$ predictions. This is to be compared to the $30$ minutes needed by Arome-NWC to calculate 24 forecasts. In both cases (Kriging and Random Forest), the calculation time is reduced. Nevertheless, Random Forest have the asset of having a very short calculation time that is more suitable for operational use. One should keep in mind that the calculation time changes with the size of the region that is being considered and the number of samples used. It is important to notice that once the surrogate model has been built, making a new prediction is very quick. We chose here to do a prediction for every minute, but the time interval can be easily reduced.

\section*{Acknowledgments}
\emph{This work is part of the Deep4Cast project\footnote{https://www.researchgate.net/project/Deep4Cast}, funded by STAE.  We thank Météo-France for providing us with meteorological data for our calculations. In particular, thank you for accepting our request and providing us with data that are usually not stored.}

%
%
\section*{Data Statement}
The data used here were kindly provided by Météo-France. If access is required, please contact Michaël Zamo (michael.zamo@meteo.fr).

\bibliographystyle{ametsoc2014}
\bibliography{references}


\end{document}